\documentclass[12pt]{article}

\usepackage{amsfonts}
\usepackage{amssymb}
\usepackage{amstext}
\usepackage{amsmath,bm}
\numberwithin{equation}{section}

\usepackage{caption}
\usepackage{graphicx}
\usepackage[all]{xy}
\usepackage{color}
\usepackage{float}

\def\be{\begin{equation}}
\def\ee{\end{equation}}
\def\ba{\begin{eqnarray}}
\def\ea{\end{eqnarray}}
\def\lb{\label}
\def\nn{\nonumber}

\def\e{\varepsilon}
\def\l{\lambda}

\def\id{\mbox{\em 1\hspace{-3.4pt}I}}

\def\Z{\mathbb Z}

\def\1{1\!\!{\rm I}}

\def\subbbc{{\rm C}\kern-3.3pt\hbox{\vrule height4.8pt width0.4pt}\,}

\begin{document}

\begin{titlepage}

\begin{flushright}
Dedicated to the memory of \\ Professor Christo Christov (1915-1990)
\end{flushright}

\vskip 1in

\begin{center}

{\Large {\bf Quantum entanglement}}

\vspace{4mm}

Ludmil Hadjiivanov and Ivan Todorov

Institute for Nuclear Research and Nuclear Energy\\
Tsarigradsko Chaussee 72, BG-1784 Sofia, Bulgaria\\
e-mails: lhadji@inrne.bas.bg, ivbortodorov@gmail.com

\end{center}

\begin{abstract}

Expository paper providing a historical survey of the gradual transformation of the ``philosophical discussions''
between Bohr, Einstein and Schr\"odinger on foundational issues in quantum mechanics into a quantitative prediction
of a new quantum effect, its experimental verification and its proposed (and loudly advertised) applications.
The basic idea of the 1935 paper of Einstein-Podolsky-Rosen (EPR) was reformulated by David Bohm for a finite
dimensional spin system. This allowed John Bell to derive his inequalities that separate the prediction of
quantum entanglement from its possible classical interpretation. We reproduce here their later (1971) version,
reviewing on the way the generalization (and mathematical derivation)
of Heisenberg's uncertainty relations (due to Weyl and Schr\"odinger) needed for the passage from EPR to Bell.
We also provide an improved derivation of the quantum theoretic violation of Bell's inequalities.
Soon after the experimental confirmation of the quantum entanglement (culminating with the work
of Alain Aspect) it was Feynman who made public the idea of a quantum computer based on the observed effect.

\end{abstract}

\tableofcontents

\end{titlepage}

\vfill\eject

\section{Introduction}

One thing that troubled Einstein most with the Copenhagen
Interpretation was the ``instantaneous reduction of the wave
function'' -- and hence of the probability distribution -- when a
measurement is performed. After Bohr's talk at the fifth (the
famous!) Solvay Congress in October 1927, he made a comment
concerning the double slit experiment. Bohr's probability wave is
spread over the detector screen, but as soon as the electron is
detected at one point, the probability becomes zero everywhere
else -- instantly (see, e.g., \cite{G} Ch. 8, Sect. {\it Berlin
and Brussels}). If the reduction of the probability wave of a
single particle may not have seemed so paradoxical, the
consequences of the thought experiment that Einstein, Podolsky and
Rosen \cite{EPR} proposed in 1935 appear really drastic. It
involves two correlated particles such that measuring the
coordinate or the momentum of one of them fixes the corresponding
quantity of the other, possibly {\it distant particle}. Only two
senior physicists reacted to the EPR paper at the time:
Schr\"odinger \cite{S35} sympathized with the authors and, after a
correspondence with Einstein (\cite{G}, Ch. 9, Sect. {\it The cat
in the box}; \, \cite{F14}, Sect. 1.3), introduced the term {\it
entanglement} (as well as the notorious cat -- \cite{S/T}); Bohr
\cite{B35} challenged Einstein's notion of physical reality and
rejected the idea that its quantum mechanical description is
incomplete. The younger ``working particle physicists'' ignored
the discussion (probably dismissing it as ``metaphysical''). A
gradual change of attitude only started in the 1950's with the
work of David Bohm, an American physicist that had to leave the US
after loosing his job as an early victim of the McCarthy era
\cite{F}. He reformulated the EPR paradox (first in his textbook
on Quantum Theory \cite{B51}, then, more thoroughly, in an article
with his student Aharonov \cite{BA}) in terms of the electron spin
variables. The reduction to a finite dimensional quantum
mechanical problem soon allowed a neat formulation -- in the hands
of John Bell \cite{B64, B} -- and opened the way to its
experimental verification. In a few more decades it gave rise to a
still hopeful and fashionable outburst of activity under the
catchy names of ``quantum computers'' or ``quantum information''.

The present paper aims to highlight the key early steps of what is
being advertised as ``a new quantum revolution'' \cite{A13}. We
begin in Sect. 2 by reviewing the post Heisenberg development and
understanding of the {\it uncertainty relations} using the algebraic formulation
of quantum theory. Sect. 3 is devoted to the early history of the subject -- from EPR,
Schr\"odinger and Bohr through Bohm to Bell, Clauser, Shimony et al. \cite{CHSH} who proposed
to use polarized photons to test Bell's inequalities to the ultimate realization of this proposal
in the work of Alain Aspect (see his later reviews \cite{A, A13} and references cited there).
Sect. 4 deals with the actual derivation of Bell-CHSH inequalities for classical
``local hidden variables'', following \cite{B71};
in describing their violation in quantum theory we introduce a maximally entangled U(2) invariant state.
Sect. 5  overviews the work of Feynman \cite{F82} and of Manin and Shore (see \cite{M} and references therein)
and ends with a general outlook. We briefly discuss the (partly philosophical -- as reviewed in \cite{F14})
issue of {\it nonlocality} siding with the dissenting view of a mathematical physicist \cite{D}.

\bigskip

\section{Weyl-Schr\"odinger's uncertainty relations}

\setcounter{equation}{0}

Much of the early discussions on the meaning of quantum mechanics, turned around
the uncertainty relations which restrict the set of legitimate questions one can ask
about the microworld. Heisenberg justified in 1927 his uncertainty principle for the
measurement of the position $x$ and the momentum $p$ by analyzing the
unavoidable disturbance of the microsystem by any experiment designed to determine
these variables. In Weyl's book \cite{W} of the following year (1928) one finds a
derivation of a more precise relation for {\it the product of dispersions (mean square
deviations)} $\sigma_x^2\sigma_p^2$ from the properties of the wave function describing
the quantum state. Soon after, Schr\"odinger \cite{S} and others (for a review and more
references -- see \cite{T}) extend Weyl's analysis to general pairs of noncommuting operators
-- a necessary step towards a realistic test of entanglement. Let us point out
that, while the validity of Heisenberg's popular arguments for the limitations concerning
individual measurements have been questioned in recent experiments \cite{R-S},
the mathematical results about the dispersions of incompatible observables which we proceed to review are impermeable.

In order to display the generality and simplicity of the uncertainty relations we shall
adopt (and begin by reminding) the algebraic formulation of quantum theory, which, having the aura
of abstract nonsense, is seldom taught to physicists.

We start with a (noncommutative) {\it unital star algebra} $\mathcal{A}$ -- a complex vector
space equipped with an associative multiplication (with a unit element 1) and an
{\it antilinear antiinvoltion} $*$ such that
\begin{equation}
\label{*}
(AB)^*=B^*A^*\,, \ \ (\lambda A)^*=\bar{\lambda}A^*\,,\ \ (A^*)^*=A \quad
\mbox{for} \quad A,B\in\mathcal{A}\,,\ \ \lambda\in\mathbb{C}\
\end{equation}
where the bar over a complex number stands for complex conjugation. The {\it hermitean} elements
$A$ of $\mathcal{A}\,$ (such that $A^*=A$) are called {\it observables}.
A {\it state} is a (complex valued) {\it linear functional} $\langle A\rangle$ on
$\mathcal{A}$ satisfying {\it positivity}: $\langle A^*A \rangle \geq 0$ and {\it normalization}:
$\langle 1 \rangle =1\,.$

\medskip

{\it Proposition 2.1}
If $A$ is an observable, $A=A^*$, then its {\it expectation value} $\langle A\rangle$ is real; moreover,
\begin{equation}
\label{AB}
A^*=A\,,\ B^*=B \quad \Rightarrow\quad \langle BA\rangle =\overline{\langle AB \rangle}\ .
\end{equation}

\smallskip

{\it Proof}.
The implication (\ref{AB}) is a consequence of the positivity (and hence the reality)
of both $\langle (A+B)^2\rangle$ and $\langle (A+iB)(A-iB)\rangle$.
The reality of $\langle A\rangle$ for $A=A^*\,$ follows from (\ref{AB}) for $B=1\,.$

\medskip

{\it Remark 2.1}
The more common definition of a ({\it pure}) quantum state as a vector $|\Psi\rangle\,$ of norm $1\,$ in a Hilbert space
(or rather a $1$-dimensional projection $|\Psi\rangle\langle \Psi|\,$) is recovered as a special case for
$\langle A \rangle = \langle \Psi|\, A\, |\Psi\rangle \equiv {\rm tr}\, (A \, |\Psi \rangle \langle \Psi | \,)\,.$
The reality of the expectation value in this case appears as a corollary of the spectral decomposition
theorem for hermitean operators while in the above formulation it is an elementary consequence of the algebraic
positivity condition. Furthermore, our definition applies as well to a {\it mixed} state (or a {\it density matrix}).
The set of all (admissible) states form a convex manifold $\mathcal{S}$. The pure states appear as {\it extreme points}
(or indecomposable elements) of $\mathcal{S}$.

\medskip

We now proceed to the formulation and the {\it elementary proof} of the Schr\"odinger's uncertainty relation which
is both stronger and more general than Weyl's precise mathematical statement of Heisenberg's principle.
Let $A, B$ be two (noncommuting) observables with expectation value zero. (The general case is reduced to
this by just replacing $A, B$ by $A - \langle A \rangle \,,\, B - \langle B\rangle\,.$)

\medskip

{\it Proposition 2.2} ({\it Schr\"odinger's uncertainty relation})
The product of the dispersions of the above observables exceeds the sum of squares of the expectation values
of the hermitean and the antihermitean parts of their product:
\begin{eqnarray}
\label{SchrUR}
&&\sigma_A^2 \sigma_B^2\ (\,\equiv \langle A^2 \rangle \langle B^2 \rangle\,)\ \geq
\langle AB \rangle \langle BA \rangle =
\nonumber \\
&&= (\frac{1}{2} \langle AB+BA \rangle)^2 + (\frac{1}{2i} \langle AB-BA \rangle)^2 \geq
|\frac{1}{2}\langle [A, B]\rangle |^2.
\end{eqnarray}

\smallskip

The {\it proof} consists of a direct application of the elementary Schwarz inequality for a positive quadratic form,
taking (\ref{AB}) into account. The Heisenberg-Weyl uncertainty relation is obtained as a special case for
$A=p\,,\, B=q$ using $[q,p] = i \hbar$. We shall apply the inequality (\ref{SchrUR}) to the case of two
orthogonal polarization vectors in Sect. 4 below.

\medskip

{\it Remark 2.2}
We note that there are further strengthenings of the uncertainty relations (see e.g. \cite{MP})
but Proposition 2.2 will be sufficient for our purposes.

\section{From EPR to Bell's inequalities}

\setcounter{equation}{0}

In 1935, already at Princeton, (the 56-year-old) Einstein with two younger collaborators,
Boris Podolsky (Taganrog, 1896 -- Cincinnati, 1966) and Nathan Rosen (Brooklyn, 1909 --
Haifa, 1995) proposed something new. They consider a state of two particles travelling with
opposite momenta (in opposite directions) along the $x$-axis with a (non-normalizable) wave function
\begin{equation}
\label{EPR}
\Psi(x_1, x_2) = \int u_p(x_1)\,u_{-p}(x_2)\, e^{i\frac{p\,d}{\hbar}}\frac{dp}{2\pi\hbar} = \delta(x_1-x_2+d) \ ,\ \
u_p(x)= e^{i\frac{px}{\hbar}}\, .
\end{equation}
If one measures the position of the first particle $x_1$ the position of the second one would be fully determined ($x_2 = x_1+d$).
If one measures instead its momentum and finds the value $p_1=p\,$ then the momentum of the second particle will be $p_2=-p$.
None of the operations on the first particle should disturb the second one (as the distance $d$ between the two can be made
arbitrarily large). It then appears that the second particle should have both definite position and definite momentum.
As quantum mechanics cannot
accommodate both the authors conclude that it has to be considered incomplete. The precise wording of the abstract of \cite{EPR} is
more nuanced: ``... A sufficient condition for the reality of a physical quantity is the possibility of predicting it with certainty,
without disturbing the system. In quantum mechanics in the case of two physical quantities described by non-commuting operators, the
knowledge of one precludes the knowledge of the other. Then either (1) the description of reality given by
the wave function in quantum mechanics is not complete or (2) these two quantities cannot have simultaneous reality...''.
(The paper was actually written by Podolsky and Einstein was not happy: he found the wording obscuring the simple message...
-- see \cite{F14}, Sect. 1.3.)

Two quantum theorists responded to the EPR paper (both born in the 1880's): Schr\"odinger \cite{S}, a sympathizer, who,
after a correspondence with Einstein, introduced the term {\it entanglement}  \cite{S/T}, and Bohr, the father of the
Copenhagen Interpretation, who was upset. Bohr's reaction was recorded by his faithful collaborator (since 1930),
the Belgian physicist L\'eon Rosenfeld (1904-1974 -- characterized in \cite{J} as a Marxist defender of complementarity)
and later told by Bohr's grandson Tomas. In Rosenfeld's words:

``This onslaught came down upon us as a bolt from the blue. ... As soon as Bohr had heard my report of
Einstein's argument, everything else was abandoned: we had to clear up
such a misunderstanding at once. We should reply by taking up the same
example and showing the right way to speak about it. In great excitement,
Bohr immediately started dictating to me the outline of such a reply. Very
soon, however, he became hesitant: 'No, this won't do, we must try all over
again ... we must make it quite clear ...' ... Eventually, he broke off with
the familiar remark that he 'must sleep on it'. The next morning he at
once took up the dictation again, and I was struck by a change of the tone:
there was no trace of the previous day's sharp expressions of dissent. As I
pointed out to him that he seemed to take a milder view of the case, he smiled: 'That's a sign', he said,
'that we are beginning to understand the problem.' Bohr's reply was that yes, nature
is actually so strange. The quantum predictions are beautifully consistent,
but we have to be very careful with what we call `physical reality'.''

Bohr's reply \cite{B35} to EPR carried the same title {\it Can quantum-mechanical description of physical reality
be considered complete?} and also appeared in the Physical Review four months later. Compared with the clear message
of EPR it appears tortured: `` Indeed the finite interaction between object and measuring agencies conditioned
by the very existence of the quantum of action entails -- because of the impossibility of controlling the reaction
of the object on the measuring instruments if these are to serve their purpose -- the necessity of a final renunciation
of the classical ideal of causality and a radical revision of our attitude towards the problem of physical reality.''
No wonder that that the younger generation was repelled by such a metaphysical twist in the discussion.

The next step was only made some 15 years later by David Bohm (1917-1992) who had the opportunity to discuss the
matter with Einstein. In his 1951 book on Quantum Theory and later in \cite{BA} he reduced the problem in the case
of electron spins for observables that are {\it dichotomic} -- a crucial advance for the subsequent development
(see \cite{A13}, Sect. 3). It was for such type of observables that Bell could derive in 1964 \cite{B64} his inequalities.
Bell's paper, which now counts over 9000 citations, remained essentially unnoticed until the 1969 work of Clauser et al.
 \cite{CHSH} (see Chapter 7 of \cite{F}, in particular, Picture 7.4).

{\it Remark 3.1} Bell was originally motivated by the work of Bohm on hidden variables and by the observation that
it provided a counterexample to von Neumann's ``no hidden variable'' theorem (of 1932 -- translated into English,
as if just to counter Bohm's theory, in 1955 \cite{vN}). Remarkably, he ended up by establishing a better theorem
of this type -- both more general and concrete -- stimulating new experiments. (For interesting later comments
on von Neumann's theorem and Bell's critique -- see \cite{R, B10}.)

Freire is right to call all participants in these developments {\it quantum dissidents} -- physicists trying to
upset what the historian of physics Max Jammer (1974) had termed the ``almost
unchallenged monocracy of the Copenhagen school'' (see Sect. 2.1 of \cite{F}). Having a doctorate
in philosophy (under Carnap in 1953) before acquiring a second doctorate in physics (under Eugene
Wigner in 1962) was not viewed as an asset for Abner Shimony in the physics community.
The authoritative intervention of Wigner (Nobel Prize, 1963) was needed to defend him from strong and
unfair criticism after his first paper on the foundations of quantum mechanics (see Sect. 7.3:
``Philosophy enters the labs: the first experiments'' of \cite{F}). When John Clauser still a
graduate student wanted to prepare an experiment to check Bell's inequalities (independently of Shimony)
and asked the advice of Feynman about his project the answer was: ``You will be wasting your time'' (see \cite{N}).
Happily, when Shimony learned about Clauser's proposal to do the same (Bell-) type experiment that he
has assigned to his graduate student Horne, he followed Wigner's advice and called Clauser; so rather
than engaging in a fierce competition they started a fruitful collaboration. Even after
their landmark 1969 paper \cite{CHSH} and the first experimental confirmation  of the quantum mechanical violation
of the Bell-CHSH inequalities (by Clauser and Freedman, 1972) debates on the
foundations of quantum mechanics were formally forbidden by the editor (Samuel Goudsmit, 1902-1978)
of Physical Review. Clauser and Shimony then started publishing {\it Epistemological Letters} -- a
hand-typed, mimeographed, ``underground'' physics newsletter about quantum physics distributed by a Swiss
foundation (1973-1984). According to Clauser, much of the early work on Bell's theorem was
published only there (including Bell's paper \cite{B75} and the responses to it by Shimony, Clauser and Horne;
for more on this story -- see \cite{K}).

\section{Inequalities separating hidden variable and quantum predictions for a 2-state system}

\setcounter{equation}{0}

The EPR paradox rephrased in terms of dichotomic  observables says
that measuring e.g. a photon polarization we shall instantly
determine the polarization of its distant entangled partner
without disturbing it. On the other hand, as we shall recall
shortly, polarizations in two directions differing by an angle
$\theta\,$ cannot be determined simultaneously unless
$\sin  2\,\theta=0\,.$ The paradox would be resolved if photon
polarization in all directions was fully determined  by some
additional statistical parameters termed {\it hidden variables}
(HV). This is a natural assumption. To cite \cite{A13} ``when
biologists observe strong correlations between some features of
identical twins they can conclude that these features are
determined by identical chromosomes. We are thus led to admit that
there is some common property whose value determines the result of
polarization. But such a property, which may differ from one pair
to another, is not taken into account by the quantum state which
is the same for all pairs. One can thus conclude with EPR that
Quantum Mechanics is not complete.'' It was Bell \cite{B64, B} who
realized that even without specifying the nature of the hidden
parameters, just assuming that they are not affected by changes in
the distant experimental arrangement, their existence implies
certain inequalities in the probability distribution of the
polarization that are violated if the quantum mechanical
predictions hold. Our survey below of this landmark work takes into account subsequent development by
Clauser et al. \cite{CHSH} and by Bell himself \cite{B71}.

\smallskip

\subsection{Bell-CHSH inequalities for (classical) hidden variables}

\begin{figure}[htb]
\centering
\includegraphics[width=1.0\textwidth]{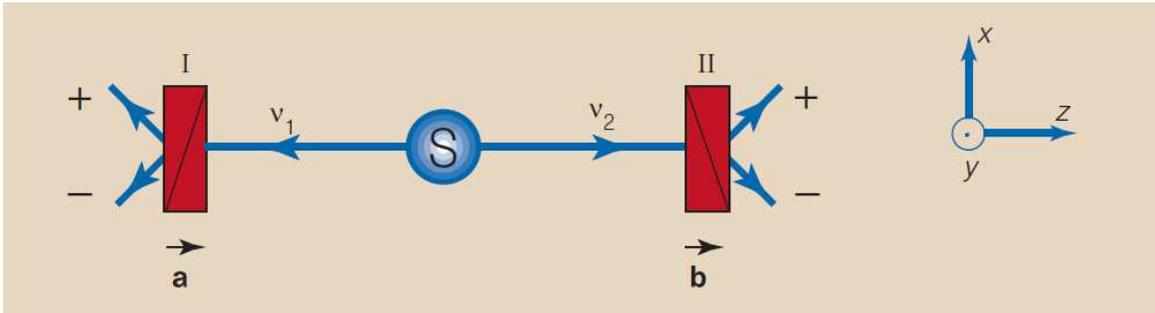}
\caption{\footnotesize{EPR-Bohm Gedanken experiment with photons \cite{A}. The two
photons travelling in opposite directions away from a source are analyzed by linear polarizers
in orientations ${\bf a}\,$ and ${\bf b}\,.$ One measures the probabilities of joint detections
in the output channels at various orientations of the polarizers.}}
\label{Fig1}
\end{figure}

One starts with a pair of linearly polarized photons emitted by a
source characterized by some supplementary parameters $\l\,,$ and
two analyzers, $A\,$ in orientation ${\bf a}\,$ and $B\,$ in
orientation ${\bf b}\,$ which may depend on some additional
parameters $\l'\,.$ The photons are assumed to have opposite
momenta; the polarizations ${\bf a}\,$ and ${\bf b}\,$ are then
represented by two unit vectors in a plane (say $(x, z)$)
orthogonal to their common line of propagation ($y$) (see Fig. 1). The possible
outcomes of the polarization measurement $A({\bf a}, \l\,)$ will
be taken $1$ for a polarization along ${\bf a}\,$ and $-1\,$ for a
polarization in the orthogonal direction in the same plane
(and similarly for $B({\bf b}, \l\,)$ and ${\bf b}\,,$ respectively).
If we consider, following \cite{B71}, averaging with respect to the
analyzers' parameters $\l'\,$ we should replace $A({\bf a}, \l)$
and $B({\bf b}, \l)$ by their mean values which satisfy
\be
-1\leq {\bar A}({\bf a}, \l\,)\leq 1\ ,\qquad -1\leq {\bar B}({\bf b}, \l\,)\leq 1\ .
\lb{ABHV}
\ee
(To quote from \cite{B71}: ``In practice, there will be some occasions on which one or both
instruments simply fail to register either way. One might then count $A\,$ and/or $B\,$ as zero
in defining ${\bar A}\,,\,{\bar B}$.'' Then (\ref{ABHV}) still holds.) Introduce further a normalized
probability measure
\be
d \,\mu (\l )\geq 0,\qquad \int
d\, \mu (\l) = 1\ .
\lb{rho}
\ee
Knowing ${\bar A}\,,\, {\bar B}\,$ (\ref{ABHV}) and the measure (\ref{rho}) one can compute the probabilities
for various outcomes. Assuming that of the measurements $A\,$ and $B\,$ are independent we deduce that the
joint probability of a pair of outcomes is equal to the product of the separate probabilities for each of them,
so that the statistical correlation function (expectation value) would be given by the {\it bounded} mean value of their product:
\be
-1\leq E ({\bf a}, {\bf b} ) = \langle\, A({\bf a}) \, B({\bf b})\, \rangle_{HV} :=
\int
d \mu (\l)\, {\bar A}({\bf a}, \l) \, {\bar B}({\bf b}, \l)\leq 1 \ .
\lb{EHV}
\ee

{\it Remark 4.1} Here we stick to the traditional notation $A, B$ (reminiscent to the "Alice and Bob" used in cryptography)
which involves a redundancy: for a fixed vector ${\bf a}, A({\bf a})=B({\bf a})$ so that we are dealing with a single
physical quantity.  We shall make this explicit in our treatment of the quantum case.

Let ${\bf a}'$ and ${\bf b}'$ be alternative directions of polarization. We shall prove the following inequality for
the sum of absolute values of particular linear combination of correlations in any classical (HV) theory:
\be
|E({\bf a}, {\bf b})-E({\bf a}, {\bf b}')| +|E({\bf a}', {\bf b})+E({\bf a}', {\bf b}')|\leq 2\ .
\lb{B71}
\ee
Indeed (\ref{B71}) holds as a consequence of the following chain of inequalities:
\ba
&&|E({\bf a}, {\bf b})-E({\bf a}, {\bf b}')| =\nn\\
&&=|\int d\mu(\l){\bar A}({\bf a},\l){\bar B}({\bf b}, \l)\, (1\pm {\bar A}({\bf a}',\l){\bar B}({\bf b}', \l))- \nn\\
&&- {\bar A}({\bf a},\l){\bar B}({\bf b}', \l)\, (1\pm {\bar A}({\bf a}',\l){\bar B}({\bf b}, \l))|\leq\nn\\
&&\leq \int d\mu(\l)\,(1\pm {\bar A}({\bf a}',\l){\bar B}({\bf b}', \l)) 
+\int d\mu(\l)\,(1\pm {\bar A}({\bf a}',\l){\bar B}({\bf b}, \l))=\nn\\
&&= 2\pm (E({\bf a}', {\bf b}')+E({\bf a}', {\bf b}))\ .
\lb{Bell1}
\ea
Introducing the linear combination of $2$-point correlation functions
\be
S ({\bf a},{\bf b}, {\bf a}', {\bf b}') =
E ({\bf a}, {\bf b} ) - E ({\bf a}, {\bf b}' ) + E ({\bf a}', {\bf b} ) + E ({\bf a}', {\bf b}' ) =
\int d \mu (\l)\,S (\l; {\bf a},{\bf b}, {\bf a}', {\bf b}')\ ,
\lb{SHV2}
\ee
we deduce from (\ref{B71}) the result first obtained (under slightly more restrictive conditions) by CHSH:
\be
|S ({\bf a},{\bf b}, {\bf a}', {\bf b}')| \le 2\ .
\lb{Bell}
\ee
We stress that the assumption of positivity of the measure $d\mu$ is essential for the validity of the Bell-CHSH
inequalities -- a point also emphasized by Feynman \cite{F82}.
Admitting "negative probabilities" one can reproduce all quantum mechanical results!

\subsection{Quantum mechanical treatment of pairs of entangled photons}

It is customary to consider a 2-dimensional Hilbert space $\mathcal{H}$ of (a single) photon polarization with a basis
$|\e \rangle\,,\ \e =\pm\,$ where $+ (-)$ corresponds to a linear polarization along the $z\, (x)$-axis, respectively.
In fact, this labeling is not quite complete. Dealing with a pair of entangled photons one should also indicate the
sign of the photon momentum $p\,$ -- along the positive or the negative $y\,$ axis. Observing that changing the sign of
$p\,$ amounts to complex conjugation of the wave function we shall put a bar over the state vector of the second photon
that moves in the opposite direction to the first. (This amounts to introducing a real structure in our Hilbert space
in terms of a linear isomorphism between the dual space $\mathcal{H}'$ of "bra vectors" and the space $\bar{\mathcal{H}}$
of complex conjugate "kets".) A maximally entangled state in the 4-dimensional complex Hilbert space
$\mathcal{H}\otimes \bar{\mathcal{H}}$ is given by
\be
\Psi = \frac{1}{\sqrt{2}}\sum_{\e=\pm}|\e\rangle\otimes\bar{|\e\rangle}\quad
(\, = u\otimes\bar{u}\,\, \Psi \quad \mbox{for} \quad u\in U(2)\,
)\ . \lb{psi} \ee It is independent of the choice of basis
$|\e\rangle$ being the unique $U(2)$-invariant pure state in
$\mathcal{H}\otimes \bar{\mathcal{H}}$. (Note that this is not
true for the customarily used real $O(2)$-invariant substitute of
(\ref{psi}).)

A state $|\theta, \e \rangle \in \mathcal{H}$ of polarization $\e$ in a direction of angle $\theta$ with
respect to the $z$-axis (in the $(z, x)$-plane) is given by:
\be
| \,\theta, \e \rangle := \cos \theta\, |\e \rangle + \e\, \sin \theta \,| -\e \rangle\ ,\qquad
\e = \pm \ .
\lb{te}
\ee
The operators $A_i(\theta)\,,\ i= 1, 2\,$ corresponding to the analyzers of the first and the second photon are given by
\ba
&&A_1(\theta)= A(\theta)\otimes 1\ , \quad A_2(\theta)= 1\otimes A(\theta)\ , \quad
A(\theta) = \cos 2\theta\, \sigma_3 + \sin 2\theta \,\sigma_1\ ,\nonumber\\
&&\sigma_3\, |\e \rangle= \e\, |\e \rangle\ , \quad \sigma_1\, |\e \rangle = |-\e \rangle\ .
\lb{Ai}
\ea
Here $A(\theta)$ is the operator with eigenvectors $|\theta, \e \rangle$ corresponding to eigenvalues $\e$:
\be
A(\theta)\,|\theta, \e \rangle = \e \,|\theta, \e \rangle\  .
\lb{A}
\ee
The $2$-point correlation function of the product $A_1 A_2\,$ in the state $\Psi\,$ (\ref{psi}) is given by
\be
E(\theta_1, \theta_2):= \langle \,\Psi|A_1(\theta_1) A_2(\theta_2) | \Psi \rangle = \cos 2(\theta_1-\theta_2)\ .
\lb{ampl}
\ee

\smallskip

{\it Remark 4.2~} This result can also be expressed as a linear combination of individual
probabilities $P_{\e_1 \e_2} (\theta_1-\theta_2)\,$ defined below (cf. \cite{A, A13}):
\ba
&&E(\theta_1, \theta_2) := \langle \Psi | A_1(\theta_1) A_2(\theta_2) | \Psi \rangle
= \sum_{\e_1, \e_2} \e_1 \e_2 \,|\langle \theta_1, \e_1|\otimes\langle \theta_2, \e_2| \Psi \rangle |^2 = \nn\\
&&= P_{++} (\theta_{12}) + P_{--} (\theta_{12}) - P_{+-} (\theta_{12}) - P_{-+} (\theta_{12})
\lb{probQM}
\ea
where
$$\theta_{ij} :=  \theta_i - \theta_j \ ,\quad
P_{\e_1 \e_2} (\theta_{12}) := |\langle \theta_1, \e_1|\otimes\langle \theta_2, \e_2| \Psi \rangle |^2\ =
\frac{1}{4}(1+\e_1 \e_2 \cos 2 \theta_{12})\ .
$$

\smallskip

{\it Remark 4.3~} Another way to express the quantum mechanical $2$-point correlation function is to use the fact that
\be
{\rm tr}_2 (A_2 (\theta_2) |\Psi \rangle \langle \Psi | ) =  A(\theta_2 ) \,\rho = \frac{1}{2}\,A(\theta_2)\ ,
\qquad \rho = \frac{1}{2} \, \id\ = \frac{1}{2}\,\sum_{\e} | \e \rangle \langle \e |
\lb{rho2}
\ee
to write
\be
E(\theta_1, \theta_2) = {\rm tr} \,( A_1(\theta_1) A_2(\theta_2) | \Psi \rangle \langle \Psi |)  =
{\rm tr} \,(A(\theta_1) A(\theta_2)\,\rho ) = \frac{1}{2}\,{\rm tr} \,(A(\theta_1) A(\theta_2))\ .
\lb{r1}
\ee
The corresponding joint probability distributions $P_{\e_1 \e_2} (\theta_{12})\,$ are given by R. Stora's formula
\be
P(a, b) = \frac{1}{2}\,\left(\langle a|b \rangle \langle b|\rho|a \rangle + \langle a|\rho|b \rangle \langle b|a \rangle \right)
\lb{Stora}
\ee
(cf. Eq. (4.7) of Sect. 4.1 of \cite{T12}) for
$|a\rangle = |\theta_1, \e_1\rangle\,,\ |b\rangle = |\theta_2, \e_2\rangle\,$ from (\ref{te}).

\smallskip

The matrix $\rho\,$ is a particular case of {\em reduced density matrix} (corresponding to a pure state of the
composite system), a notion introduced in \cite{D30}; it describes the state as viewed by
an observer attached to one of the subsystems.

\smallskip

{\it Comment.~} Any state in a $2$-dimensional Hilbert space can be realized as a density matrix
$\rho({\underline a})\,$ labeled by a vector ${\underline a}\,$ in the unit ball:
\be
\rho = \rho ({\underline a}) = \frac{1}{2}\,\left(\id + {\underline a} \cdot {\underline{\sigma}} \right) =
\frac{1}{2}\,\begin{pmatrix} 1 + a_3 & a_1 - i a_2\cr a_1 + i a_2 & 1-a_3\end{pmatrix}\ ,\quad
{\underline{a}}^2 = a_1^2 + a_2^2 + a_3^2 \le 1\ .
\lb{Bloch}
\ee
The pure states belong to the boundary of this domain, the {\em Bloch sphere}
\be
{\mathbb S}^2 \, \simeq \, {\mathbb S}^3 / {\mathbb S}^1 = SU(2)/\,U(1)\ .
\lb{BlS}
\ee
The {\em von Neumann entropy} $S = - {\rm tr}\, \rho\, \log_2 \rho\,$
varies between $0\,$ (for a pure state) and $1\,,$ for the maximally mixed state $\rho = \frac{1}{2}\,\id\,.$

\smallskip

It follows that the quantum mechanical counterpart of (\ref{SHV2}) is
\ba
&&S(\theta_1,\theta_2, \theta_3, \theta_4) =
E(\theta_1, \theta_2 ) - E(\theta_1, \theta_4 ) + E(\theta_3, \theta_2 ) + E(\theta_3, \theta_4 ) = \nn\\
&&= \cos 2 \,\theta_{12} - \cos 2 \,\theta_{14} + \cos 2 \,\theta_{23} + \cos 2 \, \theta_{34}\ .\quad
\lb{SQM}
\ea
Setting the consecutive differences among the angles equal to each other, $\theta_{ii+1} = \theta\,,\ i=1, 2, 3\, $
we find the following global extremal points of the function (\ref{SQM}) (see Fig. 2):
\be
\theta = \frac{\pi}{8}\,(2k+1)\ ,\quad k\in \Z\ .
\lb{T2}
\ee
For $\,\theta\,$ an odd multiple of $\,\frac{\pi}{8} = 22.5^o\,$
Eq. (\ref{SQM}) gives
\be
S(0,\theta, 2\theta, 3\theta) =
\left\{\begin{array}{ll}
3\,\cos \frac{\pi}{4} - \cos \,\frac{3\pi}{4} = 4 \, \frac{1}{\sqrt{2}} = 2 \sqrt{2}\ ,\quad\quad\ \ \,
\theta = \,\frac{\pi}{8}\, =\, 22.5^o\\
3\,\cos \frac{3\pi}{4} - \cos \,\frac{9\pi}{4} = - 4 \, \frac{1}{\sqrt{2}} = - 2 \sqrt{2}\ ,\quad
\theta = \frac{3\pi}{8} = 67.5^o\end{array}\right.\ \ .
\lb{extr}
\ee
At these angles the predictions for the quantum mechanical entangled two photon system violate the Bell-CHSH inequalities
(\ref{Bell}) by more than $40\%\,,$ a fact that has been confirmed by numerous precision tests (see \cite{A, A13}).


\begin{figure}[htb]
\centering
\includegraphics[width=0.6\textwidth]{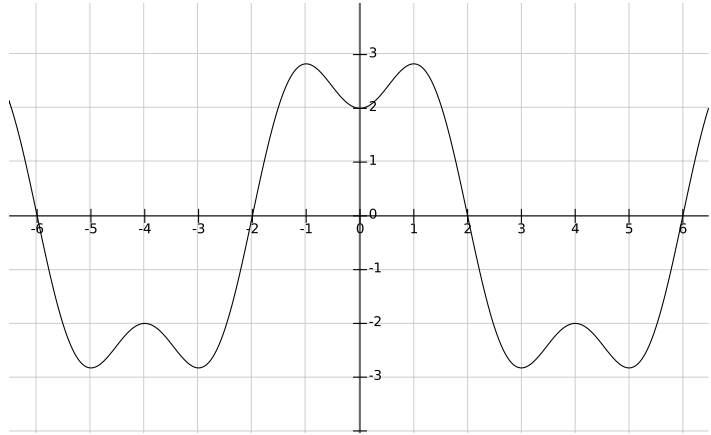}
\caption{\footnotesize{Plot of the function $3 \cos \frac{\pi}{4} x - \cos \frac{3\pi}{4} x$.}}
\label{Fig2}
\end{figure}



\section{Feynman and quantum computing. Discussion}

The story does not end with the beautiful experiments confirming
the strange predictions of quantum mechanics. In his ``keynote
talk'' at the 1st conference on Physics and Computation at MIT in
1981 Richard Feynman (1918-1988) demonstrated that he not only had
finally appreciated the work on quantum entanglement (albeit he
did not cite any names) but proposed to use it in order to
simulate quantum physics with computers \cite{F82}. During half a
century scientists were convinced that any computer is a
realization of an universal machine described in 1936 by Alan
Turing (1912-1954). Feynman noted that the behavior of entangled
photons cannot be imitated by such a classical machine and should
be used to construct a new {\it quantum computer}. If
predecessors-mathematicians (such as Manin, 1980) were interested
in new possibilities for calculations (using the ``greater capacity
of quantum states''), Feynman thinks of simulating quantum
phenomena which do not admit a classical realization in order to
better understand quantum theory: {\it we never really understood
how lousy our understanding of languages was, the theory of
grammar and all that stuff, until we tried to make a computer
which would be able to understand language} (\cite{F82} Sect. 8 p. 486).

One way or another, the cold reception of the first steps
revealing the quantum entanglement is been replaced by a hectic
activity with pretense for a new science. Books with titles like
{\it Quantum Computers and Quantum Information} are advertised by
prestigious publishers (Cambridge University Press, 2010). The
publicity is impressive, the progress is modest. A serious
mathematical result is Shor's (1994) algorithm for decomposing
large positive integers into prime factors \cite{M}. According to
it, the time needed to factor the number $N$ does not exceed a
multiple of $(\log N)^2\log\log N \log\log\log N$. It is believed
on the other hand (albeit not proven) that the time needed for a
classical factoring algorithm grows faster than any power of $\log
N$. (The problem of factoring large integers is relevant for
cryptography.) In practice, the realization of a quantum computer
is hindered by the phenomenon of {\it decoherence} in large
systems. After some twenty years of efforts (and a few billion
dollars invested) the record achieved (in 2012) by a real quantum
computer using Shor's algorithm is the factoring of $21 (=3\times 7)$.
In the words of the renowned computer scientist Leonid Levin
``The present attitude [of quantum computing researchers] is
analogous to, say, Maxwell selling the Daemon of his famous
thought experiment as a path to cheaper electricity from heat.''
(\cite{Aa}). Noticeable applications come after a long quiet
development. Quantum mechanics, created during the first quarter
of XX century is finding wide applications only after the
invention of the transistor in 1948 and the development of the
laser in the late 1950's. The true applications of the ``second
quantum revolution'' are yet to come.

If the glory of ``quantum computers'' has been overblown, the advance in our understanding and appreciation of quantum
entanglement can be hardly overstated. It had an impact even on the public awareness of the significance of quantum
theory. Here is how Jeremy Bernstein answers his question ``why people who seem
to have an aversion to more conventional science are drawn to the quantum theory?'' He believes that
\textit{the present widespread interest in the quantum theory can be traced to a single paper
with the nontransparent title ``On the Einstein-Podolsky-Rosen Paradox'', which was written in 1964 by the then
thirty-four-year-old Irish physicist John Bell. It was published in the obscure journal Physics, which expired
after a few issues.} (\cite{Ber} p. 7). ``The philosophical discussions of the old
outsiders'' (Einstein, Bohr, Schr\"odinger) lead to a new development in quantum physics. The categorical opinion,
expressed by Lev Landau (1908-1968), the leader of the Moscow school of
theoretical physics after World War II, that ``quantum mechanics was completed by 1930 and was only questioned
later by crackpots'', was shared by the majority of active physicists worldwide. In his
famous {\it Lectures on Physics}, published in 1963 Feynman writes that all the 'mystery' of Quantum Mechanics
is in the wave-particle duality and finds nothing special in the EPR situation. It took him another 20 years
(and the work of Bohm, Bell, Clauser, Shimony, Aspect) to realize that there was another quantum mystery...

The precise meaning of the violation of the Bell-CHSH inequalities
is a matter of continuing discussion. In fact, the framework of nonrelativistic quantum mechanics is not
appropriate for testing relativistic locality and causality. The
proper playground to discuss these concepts is relativistic quantum
field theory (QFT). The great majority of authors speak of ``quantum nonlocality''.
Indeed, the mere notion of a particle spin or energy-momentum in QFT requires integrating a conserved current
over an entire 3-dimensional hypersurface. Shimony recalls \cite{Sh} that Arthur Wightman (1922-2013)
asked him ``to read the paper by Einstein, Podolsky and Rosen on an argument for hidden
variables, and find out what's wrong with the argument''. Shimony did not find anything wrong in the
argument but later figured out that the EPR framework was not appropriate to test relativistic
locality. Bell \cite{B75} realized that local commutativity of quantum fields is consistent with the entanglement
(and hence with a violation of what he calls ``local causality of quantum beables''
-- but not with sending faster than light [information carrying] signals). Twelve years later it was
demonstrated \cite{SW} that {\it maximal violation Bell's inequality is generic in (local)
quantum field theory}. The continued unqualified talk of violation of locality in quantum physics provoked
S. Doplicher \cite{D} to reiterate, after another 22 years, that there is no EPR paradox in
the measurement process in local quantum field theory. His careful treatment of the subject seems to be
ignored and the discussion is still going on unconstrained (see \cite{F14, M14, BT} among many others).

\section*{Acknowledgments}
\addcontentsline{toc}{section}{Acknowledgments}

The authors' work has been supported in part by Grant DFNI T02/6 of the Bulgarian National Science Foundation.


\end{document}